\documentclass[preprint]{JHEP}
\usepackage{epsfig}
\relax

\def\be{\begin{equation}}
\def\ee{\end{equation}}
\def\bs{\begin{subequations}}
\def\es{\end{subequations}}
\newcommand{\een}{\end{subequations}}
\newcommand{\ben}{\begin{subequations}}
\newcommand{\beq}{\begin{eqalignno}}
\newcommand{\eeq}{\end{eqalignno}}

\def \gta {\mathrel{\vcenter
     {\hbox{$>$}\nointerlineskip\hbox{$\sim$}}}}

\newcommand\fverb{\setbox\pippobox=\hbox\bgroup\verb}
\newcommand\fverbdo{\egroup\medskip\noindent%
                        \fbox{\unhbox\pippobox}\ }
\newcommand\fverbit{\egroup\item[\fbox{\unhbox\pippobox}]}
\newbox\pippobox

\def \gta {\mathrel{\vcenter
     {\hbox{$>$}\nointerlineskip\hbox{$\sim$}}}}

\def\beq{\begin{equation}}
\def\eeq{\end{equation}}

\def\4R{{{}^{(4)}R}}

\def\K5{{\kappa}}
\def\K52{{\kappa^2}}

\newcommand{\tl}{\tilde t}

\newcommand{\drho}{\delta \rho}

\newcommand{\dchi}{\delta \chi}

\newcommand{\ii}{i}
\newcommand{\jj}{j}

\newcommand{\da}{\dot{a}}
\newcommand{\db}{\dot{b}}
\newcommand{\dn}{\dot{n}}
\newcommand{\dda}{\ddot{a}}
\newcommand{\ddb}{\ddot{b}}

\newcommand{\pa}{a^{\prime}}
\newcommand{\pb}{b^{\prime}}
\newcommand{\pn}{n^{\prime}}
\newcommand{\ppa}{a^{\prime \prime}}

\newcommand{\ppn}{n^{\prime \prime}}
\newcommand{\fda}{\frac{\da}{a}}
\newcommand{\fdb}{\frac{\db}{b}}
\newcommand{\fdn}{\frac{\dn}{n}}
\newcommand{\fdda}{\frac{\dda}{a}}
\newcommand{\fddb}{\frac{\ddb}{b}}

\newcommand{\fpa}{\frac{\pa}{a}}
\newcommand{\fpb}{\frac{\pb}{b}}
\newcommand{\fpn}{\frac{\pn}{n}}
\newcommand{\fppa}{\frac{\ppa}{a}}

\newcommand{\fppn}{\frac{\ppn}{n}}

\title{Cosmological evolution with brane-bulk energy exchange}

\author{E. Kiritsis $^{1,2,4}$, G. Kofinas $^{1}$, N. Tetradis $^{3}$, T.N.
Tomaras $^{1,4}$ and V. Zarikas $^{4}$ \\

$^{(1)}$ Department of Physics and Institute of Plasma Physics, University of
Crete, 71003 Heraklion, GREECE\\
$^{(2)}$ Laboratoire de Physique Th$\acute{e}$orique de l'Ecole Normale
Sup$\acute{e}$rieure\\
24 rue Lhomond, Paris, CEDEX 05, F-75231, FRANCE\\
$^{(3)}$ Department of Physics, University of Athens, 15771 Zographou, GREECE\\
$^{(4)}$ Foundation of Research and Technology, Hellas, 71110 Heraklion, GREECE\\

{\tt E-mail: kiritsis@physics.uoc.gr, kofinas@physics.uoc.gr,
tetradis@cc.uoa.gr, tomaras@physics.uoc.gr, vzarikas@ics.forth.gr}}

 \preprint{LPTENS-02-55\\\hepth{0207060}}

\abstract{
The consequences for the brane cosmological evolution of energy exchange
between the brane and the bulk
are analyzed in detail, in the context of a non-factorizable background
geometry with vanishing effective cosmological constant on the brane.
A rich variety of brane cosmologies is obtained, depending on
the precise mechanism of energy transfer,
the equation of state of brane-matter
and the spatial topology. An accelerating era is generically a feature
of our solutions.
In the case of low-density flat universe more dark matter than in the conventional FRW
picture is predicted. Spatially compact solutions are found to delay
their re-collapse.
}

\begin{document}

\section{Introduction}

The idea that we might be living inside a defect, embedded in a higher dimensional space
has already a long history. Concerning the nature of the defect,
a solitonic codimension one or higher topological object was proposed \cite{rubakov1}
in the context of an ordinary higher dimensional
gauge field theory, coupled \cite{shaposhnikov} or not to gravity. It was soon realized, however, that,
in contrast to scalar and spin-1/2
fields, it would be difficult to confine gauge fields on such an object. Various interesting ideas
and scenaria were studied \cite{tetradis}, which have not yet given a fully satisfactory picture.
In connection with the topology and the size of the bulk space on the other hand,
the popular choice was that the extra dimensions are compact, with size
of order ${\cal O}(M_{Pl}^{-1})$.
It was argued in \cite{ABLT}, that in the context of the heterotic string
with supersymmetry broken \`a la Scherk-Schwarz this should not be true anymore.
The scale of supersymmetry breaking is tied to the
size of internal dimensions, and a desirable supersymmetry breaking scale of a few TeV
implies an extra dimension of about $10^{-16}$cm. Despite difficulties
to build a realistic model based on these ideas, the scenario was taken seriously and
analyzed further for its phenomenological consequences \cite{benakli}.

The situation is drastically different in the context of type-I
string theory \cite{lykken-witten}. A few developments led to an exciting possibility and renewed
interest in the whole idea.
First, with the discovery of the D-branes as an essential part of the
``spectrum'' in type-I  string theory, one could conjecture that we inhabit
such a D-brane embedded in a ten-dimensional bulk. The usual solitonic
defect of field theory was thus replaced by an appropriate collection
of D-branes, which by construction confine the gauge fields \cite{polchinski}, together with
all the ingredients of the standard model. All known matter and forces lie
on our brane world \cite{AKT,AADD}, with the exception of gravity,
which acts in the bulk space as well.
It was, however, pointed out \cite{ADD}, that for Kaluza-Klein
extra dimensions the gravitational force on the brane
was consistent with all laboratory and astrophysical
experimental data, as long as the extra dimensions were smaller than a characteristic scale.
This led to the exciting possibility of two extra dimensions in the sub-millimeter range.
Furthermore, it was demonstrated in the context of an appropriate effective five-dimensional
theory of gravity, that once we take into account the back reaction of the brane energy-momentum
onto the geometry of space-time,
the graviton is effectively confined on the brane and
Newton's law is reproduced to an excellent accuracy at large distances,
even with a non-compact extra
dimension \cite{rs}.

At the same time, the analysis of the cosmological consequences of the above hypotheses
attracted considerable interest. The first step was taken in \cite{binetruy}, where the
evolution of perfect fluid matter on the brane was studied, with no reference to the bulk
dynamics and, therefore, no energy transfer between the brane and the bulk.
Alternatively, a bulk-based point of view was adopted in \cite{ida}, where the cosmology
induced on a moving 3-brane in
a static Schwarzschild-AdS$_5$ background was studied, and a general interpretation of
cosmology on moving branes, together with the idea of "mirage" cosmology, were
presented in \cite{mirage}.
The equivalence of the two approaches was
demonstrated in \cite{muko}. It was also pointed out that branes provide natural mechanisms
for a varying speed of light \cite{tym}.

Energy-exchange between the brane and the bulk should in principle
be included in any realistic
cosmological scenario, and its effects have been studied in detail in the context of flat compact
extra dimensions \cite{hall}. The role of energy-exchange on brane
cosmology in the case of non-factorizable extra dimensions
has not yet been investigated extensively \cite{hebecker,ktt}, even though
the importance of energy outflow from the
brane has already been demonstrated in the context of a
Randall-
Sundrum configuration
with additional gravity induced on the brane \cite{ktt}.

The present paper is an attempt towards a more complete analysis of the cosmic evolution
of the brane in the presence of energy flow into or from the bulk.
Our aim is to generalize the picture in bulk AdS (bulk gravity plus cosmological constant)
 by considering a general bulk theory (that includes gravity).
There may be more bulk fields and more general bulk-brane couplings.
Although this can be formulated by the standard action principle and the relevant exact equations derived and studied
we have opted in this paper for a short-cut. We are analyzing the regime where the bulk energy, at the brane position,
can be consistently
neglected from the equations. Moreover, we parameterize appropriately the energy exchange as a specific power of
the matter density of the brane. Although there are other valid parameterizations, we opted for this one, motivated
by our previous work in  \cite{ktt}  where we analyzed the out-flow
of energy from the brane due to graviton radiation in the presence of induced gravity. Standard cross section calculations
give an out-flow that is a function of the temperature and other fundamental constants of the theory.
If one re-expresses the temperature in terms of the running density using the cosmological equations we obtain
a rate of flow that is a power of the density, with a dimension-full coefficient that depends on fundamental constants
and initial energy densities. This is valid typically for a whole era, that is, piecewise in the cosmological
evolution.

In a given theory, with a specific bulk content and brane-bulk couplings,
the exponent of the density as well as the coefficient
are calculable functions of the coupling constants of the Lagrangian as well as initial densities.
It is also obvious that if the bulk theory is approximately conformal the most general form of rate of outflow
 will be polynomial in the density.
Our analysis is
``phenomenological" and we parameterize different theories with different forms for the energy in/out-flow.

The paper is
organized in five sections of which this introduction is the first. In section 2 the
framework of our work is described and the approximations on the brane-bulk exchange
are presented. The effective equations for the analysis of the brane cosmology are
derived. Section 3 contains several interesting characteristic solutions of the
brane cosmology, while an exact treatment of the influx/outflow equations, relevant for
a wide range of potentially realistic
applications, is presented in section 4. Our results are
summarized and the prospects for further research along these lines are discussed in the final section.

\section{The model}

We shall be interested in the model described by the action
\be
S=\int d^5x~ \sqrt{-g} \left( M^3 R -\Lambda +{\cal L}_B^{mat}\right)
+\int d^4 x\sqrt{-\hat g} \,\left( -V+{\cal L}_b^{mat} \right),
\label{001}
\ee
where $R$ is the curvature scalar of the five-dimensional metric
$g_{AB}, A,B=0,1,2,3,5$,
$\Lambda$ is the bulk cosmological constant, and
${\hat g}_{\alpha \beta}$, with $\alpha,\beta=0,1,2,3$,
is the induced metric on the 3-brane.
We identify
$(x,z)$ with $(x,-z)$, where $z\equiv x_5$. However, following the conventions
of \cite{rs} we extend the bulk integration over the entire interval
$(-\infty,\infty)$.
The quantity $V$ includes the brane tension as well as
quantum contributions to the
four-dimensional cosmological constant.

We consider an ansatz for the metric of the form
\begin{equation}
ds^{2}=-n^{2}(t,z) dt^{2}+a^{2}(t,z)\gamma_{ij}dx^{i}dx^{j}
+b^{2}(t,z)dz^{2},
\label{metric}
\end{equation}
where $\gamma_{ij}$ is a maximally symmetric 3-dimensional metric.
We use $k=-1,0,1$ to parameterize the spatial curvature.

The non-zero components of the five-dimensional Einstein tensor are
\begin{eqnarray}
{G}_{00} &=& 3\left\{ \fda \left( \fda+ \fdb \right) - \frac{n^2}{b^2}
\left(\fppa + \fpa \left( \fpa - \fpb \right) \right) + k \frac{n^2}{a^2} \right\},
\label{ein00} \\
 {G}_{\ii\jj} &=&
\frac{a^2}{b^2} \gamma_{ij}\left\{\fpa
\left(\fpa+2\fpn\right)-\fpb\left(\fpn+2\fpa\right)
+2\fppa+\fppn\right\}
\nonumber \\
& &+\frac{a^2}{n^2} \gamma_{ij} \left\{ \fda \left(-\fda+2\fdn\right)-2\fdda
+ \fdb \left(-2\fda + \fdn \right) - \fddb \right\} -k \gamma_{ij},
\label{einij} \\
{G}_{05} &=&  3\left(\fpn \fda + \fpa \fdb - \frac{\dot{a}^{\prime}}{a}
 \right),
\label{ein05} \\
{G}_{55} &=& 3\left\{ \fpa \left(\fpa+\fpn \right) - \frac{b^2}{n^2}
\left(\fda \left(\fda-\fdn \right) + \fdda\right) - k \frac{b^2}{a^2}\right\}.
\label{ein55}
\end{eqnarray}
Primes indicate derivatives with respect to
$z$, while dots derivatives with respect to $t$.

The five-dimensional Einstein equations take the usual form
\beq
G_{AC}
= \frac{1}{2 M^3} T_{AC} \;,
\label{einstein}
\eeq
where $T_{AC}$ denotes the total energy-momentum tensor.

Assuming a perfect fluid on the brane and, possibly an additional energy-momentum
$T^A_C|_{m,B}$ in the bulk, we write
\begin{eqnarray}
T^A_{~C}&=&
\left. T^A_{~C}\right|_{{\rm v},b}
+\left. T^A_{~C}\right|_{m,b}
+\left. T^A_{~C}\right|_{{\rm v},B}
+\left. T^A_{~C}\right|_{m,B}
\label{tmn1} \\
\left. T^A_{~C}\right|_{{\rm v},b}&=&
\frac{\delta(z)}{b}{\rm diag}(-V,-V,-V,-V,0)
\label{tmn2} \\
\left. T^A_{~C}\right|_{{\rm v},B}&=&
{\rm diag}(-\Lambda,-\Lambda,-\Lambda,-\Lambda,-\Lambda)
\label{tmn3} \\
\left. T^A_{~C}\right|_{{\rm m},b}&=&
\frac{\delta(z)}{b}{\rm diag}(-\rho,p,p,p,0),
\label{tmn4}
\end{eqnarray}
where $\rho$ and $p$ are the energy density and pressure on the brane, respectively.
The
behavior of $T^A_C|_{m,B}$ is in general complicated in the presence
of flows, but we do not have to specify it further at this point.

We wish to solve the Einstein equations at the location
of the brane. We indicate by the subscript o the value of
various quantities on the brane.
Integrating equations (\ref{ein00}), (\ref{einij})
with respect to $z$ around $z=0$ gives the known
jump conditions
\begin{eqnarray}
a_{o^+}'=-a_{o^-}'  &=& -\frac{1}{12M^3} b_o a_o \left( V +\rho \right)
\label{ap0} \\
n'_{o^+}=-n_{o^-}' &=&  \frac{1}{12M^3} b_o n_o \left(- V +2\rho +3 p\right).
\label{np0}
\end{eqnarray}

The other two Einstein equations (\ref{ein05}), (\ref{ein55})
give
\begin{equation}
\frac{n'_o}{n_o}\frac{\dot a_o}{a_o}
+\frac{a'_o}{a_o}\frac{\dot b_o}{b_o}
-\frac{\dot a'_o}{a_o} =
\frac{1}{6M^3}T_{05}
\label{la1}
\end{equation}
\begin{equation}
\frac{a'_o}{a_o}\left(
\frac{a'_o}{a_o}+\frac{n'_o}{n_o}\right)
-\frac{b^2_o}{n^2_o}\left(
\frac{\dot a_o}{a_o} \left( \frac{\dot a_o}{a_o}-\frac{\dot n_o}{n_o}\right)
+\frac{\ddot a_o}{a_o}\right)
-k\frac{b^2_o}{a^2_o} =-\frac{1}{6M^3}\Lambda b^2_o
+ \frac{1}{6M^3}T_{55},
\label{la2}
\end{equation}
where $T_{05}, T_{55}$ are the $05$ and $55$ components of $T_{AC}|_{m,B}$
evaluated on the brane.
Substituting (\ref{ap0}), (\ref{np0})
in equations (\ref{la1}), (\ref{la2}) one obtains
\begin{equation}
\dot \rho + 3 \frac{\dot a_o}{a_o} (\rho + p)
= -\frac{2n^2_o}{b_o}
T^0_{~5}
\label{la3}
\end{equation}
\begin{eqnarray}
\frac{1}{n^2_o} \Biggl(
\frac{\ddot a_o}{a_o}
+\left( \frac{\dot a_o}{a_o} \right)^2
&-&\frac{\dot a_o}{a_o}\frac{\dot n_o}{n_o}\Biggr)
+\frac{k}{a^2_o}
=\frac{1}{6M^3} \Bigl(\Lambda + \frac{1}{12M^3} V^2
\Bigr)
\nonumber \\
&-&\frac{1}{144 M^6} \left(
V (3p-\rho ) + \rho (3p +\rho)
\right)
- \frac{1}{6M^3}T^5_{~5}.
\label{la4}
\end{eqnarray}

We are interested in a model that reduces to the Randall-Sundrum
vacuum \cite{rs} in the absence of matter. In this case, the first
term on the right hand side of equation (\ref{la4}) vanishes. A new scale $k_{RS}$
is defined through the relations
$V=-\Lambda/k_{RS}=12M^3 k_{RS}$.

In order to derive a solution
that is largely independent of the bulk dynamics,
the $T^5_{~5}$ term on the
right hand side of the same equation must be
negligible relative to the second one.
This is possible if we assume that the diagonal elements of the various
contributions to the energy-momentum tensor satisfy the schematic
inequality \footnote{Strictly speaking, the left hand side of (\ref{vacdom}) concerns
only the 55 components of the bulk contributions to the energy-momentum
tensor. The other components do not appear in equations (\ref{la3}), (\ref{la4})
and do not affect the cosmological evolution on the brane.}
\be
\left|
\frac{\left. T\right|^{\rm diag}_{{\rm m},B}}{
\left. T\right|^{\rm diag}_{{\rm v},B}}
\right|
\ll
\left|
\frac{\left. T\right|^{\rm diag}_{{\rm m},b}}{
\left. T\right|^{\rm diag}_{{\rm v},b}}
\right|.
\label{vacdom} \ee
Our assumption is that the bulk matter
relative to the bulk vacuum energy is much less important than
the brane matter relative to the brane vacuum energy. In this case
the bulk is largely unperturbed by the exchange of energy with the brane.
When the off-diagonal term $T^0_{~5}$ is of the same order of magnitude or
smaller than the diagonal ones, the inequality (\ref{vacdom}) implies
$T\ll \rho k_{RS}$.

At this point we find it convenient to employ a coordinate frame in
which $b_o=n_o=1$ in the above equations. This can be achieved by using Gauss
normal coordinates with $b(t,z)=1$, and by going to the temporal gauge on the
brane with $n_o=1.$ The assumptions for the form of the energy-momentum
tensor are then specific to this frame \footnote{
If the vacuum energy
dominates over the matter content of the bulk, we expect that the form of the
metric will be close to the Randall-Sundrum solution with a static bulk.
Thus, we expect (even though we cannot
demonstrate explicitly without a full solution in the bulk)
that in a generic frame, in which
\be
\left|
\frac{\left. T\right|^{\rm diag}_{{\rm m},B}}{
\left. T\right|^{\rm diag}_{{\rm v},B}}
\right|
\ll 1
\label{vacdomm}
\ee
we shall have $\dot b\simeq 0$.
Then the transformation that sets $b = 1$
is not expected to modify significantly the energy-momentum tensor.}

Using $\beta\equiv M^{-6}/144$ and $\gamma\equiv V M^{-6}/144$,
and omitting the subscript o for convenience in the following, we rewrite
equations (\ref{la3}) and (\ref{la4}) in the equivalent form
\be
\dot\rho+3(1+w)\,{{\dot a}\over a} \, \rho = -T
\label{rho}
\ee
\be
{{{\dot a}^2}\over {a^2}}=\beta\rho^2+2\gamma \rho -
{k\over{a^2}}+\chi+\lambda
\label{a}
\ee
\be
\dot\chi+4\,{{\dot a}\over
a}\,\chi=2\beta\left(\rho+{\gamma\over\beta}\right)T,
\label{chi}
\ee
where $p=w\rho$,
$T=2T^0_{~5}$ is the discontinuity of the zero-five component of the bulk
energy-momentum tensor,
and $\lambda=(\Lambda+V^2/12M^3)/12M^3$ the effective
cosmological constant on the brane.

In the equations above, Eq. (\ref{a}) is the $definition$ of the auxiliary density $\chi$.
With this definition the other two equations are equivalent to the original system
(\ref{la3},\ref{la4}).
As we will see later on, in the special case of no-exchange ($T=0$) $\chi$ represents
the mirage radiation reflecting the non-zero Weyl tensor of the bulk.

The second order equation (\ref{la4}) for the scale factor becomes
\be
{{\ddot a}\over a}=-(2+3w)\beta\rho^2-(1+3w)\gamma\rho-\chi+\lambda.
\label{decel}
\ee
As mentioned above, in the Randall-Sundrum model
the effective cosmological constant $\lambda$ vanishes, and this is the value we shall
assume in the rest of the paper.

In the special case of $w=1/3$ one may define a new function $\tilde \chi \equiv \chi+2\gamma\rho$.
The functions $\tilde \chi, \rho$ and $a$ satisfy equations (\ref{rho}) to (\ref{decel})
with $\tilde\chi$ in place of $\chi$ and $\gamma=0$. This should be expected, since for
$w=1/3$ there is no $\gamma$ left in equation (\ref{la4}).

\section{Special solutions}

Before presenting exact solutions of equations (\ref{rho})--(\ref{chi}), it
is instructive to consider a few special cases whose physical content
is more transparent.

We concentrate on the low-density region, in which $\rho \ll \gamma/\beta$. In this
case we may ignore the term $\beta\rho^2$ in the above equations compared to $\gamma\rho$.
As a result,
equations (\ref{rho})--(\ref{chi}) can be written as
\be
\dot{\rho}+3(1+w)H\rho= -T
\label{onea}
\ee
\be
H^2=\left(\frac{\dot a}{a} \right)^2 =
2\gamma\rho + \chi -\frac{k}{a^{2}}
\label{twoa}
\ee
\be
\dot{\chi} + 4 H \chi = 2\gamma T.
\label{threea}
\ee
The cosmological evolution
is determined by three initial parameters ($\rho_i$, $a_i$, $\chi_i$,
or alternatively $\rho_i$, $a_i$, ${\dot a}_i$), instead of the two
($\rho_i$, $a_i$) in conventional cosmology.
The reason is that the generalized Friedmann equation (\ref{a}) (or
\ref{twoa})
is not a first integral of the Einstein equations because of the
possible energy exchange between the brane and the bulk.

\subsection{``Mirage'' radiation for energy outflow}

Let us consider first the case
$T>0$, for which there is flow of energy out of the brane.
If the brane matter is radiation dominated with
$p=\rho/3$, equations (\ref{onea}), (\ref{threea}) have an exact solution
independently of the explicit form of $T$:
\be
\rho +\frac{\chi}{2\gamma}  =
\left(\rho_i + \frac{\chi_i}{2\gamma}\right) \frac{a_i^4}{a^4}
\label{four} \ee
and
\be
H^2=
\left(2\gamma \rho_i +\chi_i\right)
\frac{a_i^4}{a^4}-\frac{k}{a^2}.  \label{twoaa} \ee
Assume that initially $\chi_i=0$. It is clear that the effect of the
radiation on the expansion
does not disappear even if it decays during the cosmological
evolution: the Hubble parameter of equation (\ref{twoaa}) is determined by the
initial value of the energy density, diluted by the expansion in
a radiation dominated universe.
The real radiation energy density $\rho$, however, falls with time
faster than $a^{-4}$.

As a simple example we consider $T=A\rho$ with $A>0$. Then equation (\ref{onea})
can be integrated, with the result
\be
\rho=\rho_i \left({a_i \over a} \right)^{3(1+w)} e^{-At},
\label{radioactive}
\ee
where we have considered the general case $p=w \rho$
\footnote{Having neglected the term $\beta\rho^2$ from equations (\ref{twoa}) and
(\ref{threea}), these solutions are valid only if $\beta{\rho}_i^2\ll H_i^2$.
As we shall show in subsection (4.2) this condition is eventually satisfied
for all solutions in the case of outflow of the form discussed here and $k=0$.}.
From equation (\ref{four}) with $\chi_i=0$, $w=1/3$ we obtain
\be
\chi =
2 \gamma \rho_i  \frac{a_i^4}{a^4} \left( 1-e^{-At} \right).
\label{foura}
\ee
The Hubble parameter, given by
equation (\ref{twoaa}) with $\chi_i=0$, corresponds
to an initial radiation density $\rho_i$, further diluted only by
the expansion.
At late times the expansion is the consequence of
a ``mirage'' effect, arising from the
original radiation through $\chi$.

The presence of a ``mirage'' term is
possible even without energy flow. More specifically, equation (\ref{threea}) has
a solution $\chi=C/a^4$ even for $T=0$,
and $\chi$ can act as ``mirage''
radiation. The novel feature for $T\not=0$ is that
the ``mirage'' effect appears through the decay of real brane matter, even
if it was absent in the beginning.

The effect persists even if the decaying matter is not radiation.
Consider a more realistic case, in which
the brane matter consists of non-relativistic particles with
$p=0$, that can decay spontaneously into bulk matter. In this case
our previous assumption $T=A\rho$   becomes realistic.
The brane energy density is given by equation (\ref{radioactive}) with $w=0$.
For $\chi_i=0$ we find
\be
\chi =
\left[ 2\gamma A \int_{0}^t \frac{a}{a_i} e^{-At} dt \right]
\rho_i  \frac{a^4_i}{a^4}.
\label{five} \ee
For $t \to \infty$, if the scale factor grows more slowly than an exponential,
the integral converges and $\chi  \sim 1/a^4$.
Again, there is a ``mirage'' effect corresponding to an initial energy density
proportional to $\rho_i$.

It can be checked easily that,
if the decaying matter is only one component of the matter on
the brane, the ``mirage'' effect acts as an additional radiation component
at late times. Whether it can affect the density perturbations of real
matter and act as hot dark matter is an open question.

For the solution (\ref{four}),
the quantity $q\equiv{\ddot{a}}/{a}$ equals
$-(2\gamma\rho_i+\chi_i)a^4_i/a^4\equiv-\sigma/a^4$.
For $k \geq 0$, equation (\ref{twoaa}) requires
$\sigma > 0$ and, therefore, $q<0$. For $k=-1$, $\sigma$ may be negative,
leading to a loitering universe. The radiation case is not realistic
for the eternal stage  of the universe, but we are going to see in the
following a situation with  $k=-1$, $w=0$ and eternal acceleration.

\subsection{Accelerating solutions}

An interesting feature of this framework is the possible presence of
accelerating cosmological solutions. We can look for
exponential expansion with a constant Hubble parameter $H$, even if the
brane content is not pure vacuum energy.
This implies that
equations (\ref{onea})--(\ref{threea}) must have a time-independent
solution, without necessarily requiring $w=-1$.
The possible fixed points (denoted by $*$) of these equations for
$k=0$ satisfy
\begin{eqnarray}
3H(1+w)\rho_* &=& -T(\rho_*)
\label{fp1} \\
H^2_* &=& 2 \gamma \rho_* + \chi_*
\label{fp2} \\
2 H_* \chi_* &=& \gamma T(\rho_*).
\label{fp3}
\end{eqnarray}
It is clear from equation (\ref{fp1}) that, for positive matter density
on the brane ($\rho >0$), flow of energy into the brane
($T(\rho)<0$) is necessary.

The accretion of energy from the bulk depends on the
dynamical mechanism that localizes particles on the brane.
Its details are outside the scope of our discussion. However, it is
not difficult to imagine scenaria that would lead to accretion.
If the brane initially has very low energy density,
energy can by transferred onto it by
bulk particles such as gravitons.
An equilibrium is expected to set in
if the brane energy density reaches a limiting value. As a result,
a physically  motivated behavior for the function
$T(\rho)$ is to be negative for small $\rho$ and cross zero towards positive
values for larger densities.
In the case of accretion it is also natural to expect that the energy
transfer approaches a negative constant value for $\rho \to 0$.

The solution of equations (\ref{fp1})--(\ref{fp3})
satisfies
\begin{eqnarray}
T(\rho_*) &=& -\frac{3\sqrt{\gamma}}{\sqrt{2}} (1+w)(1-3w)^{1/2}\rho_*^{3/2}
\label{sol1} \\
H_*^2 &=& \frac{1-3w}{2} \gamma \rho_*
\label{sol2} \\
\chi_* &=& - \frac{3(1+w)}{2} \gamma \rho_*.
\label{sol3}
\end{eqnarray}
For a general form of
$T(\rho)$ equation (\ref{sol1}) is an algebraic equation with
a discrete number of roots.
For any value of $w$ in the region
$-1<w<1/3$ a solution is possible.
The corresponding cosmological model has a scale factor that
grows exponentially with time. The energy density on the brane remains
constant due to the energy flow from the bulk.
Our model is very similar to the steady state model of cosmology
\cite{steady}. The main
differences are that the energy density is not spontaneously
generated, and the Hubble parameter receives an additional contribution from
the ``mirage'' field $\chi$ (see equation (\ref{fp2})).

The stability of the fixed point
(\ref{fp1})--(\ref{fp3}) determines whether the exponentially expanding
solution is an attractor of neighboring cosmological flows. If we
consider a small homogeneous perturbation around the fixed point
($\rho=\rho_*+\drho$, $\chi=\chi_*+\dchi$) we find that $\drho,\dchi$ satisfy
\be
\frac{d}{dt}
\left(
\begin{array}{c}
\drho \\ \dchi
\end{array}
\right)
=
\frac{T(\rho_*)}{\rho_*}
{\cal M}
\left(
\begin{array}{c}
\drho \\ \dchi
\end{array}
\right),
\label{pert} \ee
where
\begin{eqnarray}
{\cal M}&=&
\left(
\begin{array}{cc}
-\tilde\nu +3(1-w)/(1-3w)&~~~~~ 1/\gamma(1-3w) \\
2\gamma (\tilde\nu-2/(1-3w)) &~~~~~
-2(1+9w)/[3(1+w)(1-3w)]  \end{array}
\right)
\label{mat} \\
\tilde\nu &=& \frac{d\ln |T|}{d\ln\rho} \left(\rho_* \right),
\label{aaa}
\end{eqnarray}
and we have employed the relations (\ref{fp1})--(\ref{sol3}).
If the energy flow has a simple form $T(\rho) \propto \rho^\nu$, we
have $\tilde\nu=\nu$.
The eigenvalues of the matrix $\cal M$ are
\be
M_{1,2}=\frac{
7+3w-3\tilde\nu (1+w)\pm\sqrt{
24(-3+2\tilde\nu)(1+w)+\left[7+3w-3\tilde\nu (1+w)\right]^2
}}{6(1+w)}.
\label{eigen} \ee
For $-1<w<1/3$, $0\leq\tilde\nu < 3/2$ they both have a positive real part. As we
have assumed  $T(\rho)<0$, the fixed point is stable in this case. We have
verified  this result by integrating equations (\ref{onea})--(\ref{threea})
numerically.
The approach to the fixed-point values depends on the sign of the quantity
under the square root. If this is negative the energy density oscillates
with diminishing amplitude around its fixed-point value.

For $w=-1$ we get the
standard inflation only for a value $\rho_*$ that is a zero of $T(\rho)$.
In this case there is no flow along the fifth dimension and also
$\chi_*=0$.

\subsection{The case of radiation for energy influx}

In the case of radiation the general solution of
equations (\ref{onea})--(\ref{threea})
was derived in
the previous subsection and is given by
equations (\ref{four}), (\ref{twoaa}).
The expansion is that of a radiation-dominated universe with
constant energy $(\chi_i/(2\gamma) + \rho_i)a_i^4$ per co-moving volume.
The ``mirage'' energy density is diluted $\sim a^{-4}$.

The explicit dependence on time will be discussed next in the case
of flat space ($k=0$), in which the energy density satisfies
\be
\frac{d\rho}{dt} + \frac{2}{t} \rho = -T(\rho).
\label{realen}
\ee
If $T(\rho)<0$ for all $\rho$,
and the ``friction'' term in the left hand side becomes suppressed for $t\to \infty$,
we expect an unbounded increase of $\rho$ in this limit. For $\rho \gta
\gamma/\beta$ the low energy approximation employed in this section breaks
down. The full treatment necessary in this case will be given in the next
section.

The actual situation is rather complicated and the  details
depend crucially on  the form of $T(\rho)$. Assuming that
$T(\rho)=A\rho^\nu$ with $A<0$, the exact solution of equation (\ref{realen}) for
$\nu\not= 1,3/2$ is
\be
\left(\rho \tl^2 \right)^{1-\nu}=
\left(\rho_i \tl_i^2 \right)^{1-\nu} + \frac{1-\nu}{3-2\nu}
\left(\tl^{3-2\nu}-\tl_i^{3-2\nu} \right),
\label{solrad}
\ee
where $\tl=|A|t$. For $\nu=1$ the solution is
\be
\rho=\rho_i \frac{\tl_i^2}{\tl^2} e^{\tl-\tl_i},
\label{solrad1} \ee
and for $\nu=3/2$
\be
\left(\rho \tl^2 \right)^{-1/2}=
\left(\rho_i \tl_i^2 \right)^{-1/2}
-\frac{1}{2}\ln\left(\frac{\tl}{\tl_i} \right).
\label{solrad2} \ee

For $0\leq \nu <1$ we have $\rho\sim \tl^{1/(1-\nu)}$ for $\tl\to \infty$.
For $\nu=1$ the increase of the energy density at large $\tl$
is exponential moderated by a power.
For $1<\nu<3/2$ the energy density diverges at a finite time
\be
\tl^{3-2\nu}_d=\tl^{3-2\nu}_i+\frac{3-2\nu}{1-\nu}\left(\rho_i \tl_i^2
\right)^{1-\nu}.  \label{div} \ee
A similar divergence appears for $\nu=3/2$. For $\nu>3/2$
a divergence occurs if the quantity
\be
D=\left( \rho_i \tl_i ^2 \right)^{1-\nu} -
\frac{\nu-1}{2\nu-3} \tl_i^{2\nu-3}
\label{quant} \ee
is negative. In the opposite case
$\rho \tl^2 \to 1/D$ for $t\to \infty$, and the
energy density diminishes: $\rho \sim t^{-2}$.

As we discussed earlier, it is physically reasonable
that the energy influx should stop at a certain value $\rho_{cr}$, and
be reversed for larger energy densities.
The dynamical mechanism that localizes particles on the brane cannot
operate for arbitrarily large energy densities. This modifies the
solutions above that predict an unbounded increase of
the energy density.

A final observation that will be encountered again in the next section
is that, despite of the fact that the energy density in most cases
increases for large
times, it can decrease at the initial stages.
This is obvious from eq. (\ref{realen}).
If at the time $t_r$ that the brane enters a radiation dominated era
$|T(\rho)|/\rho<2/t$, the energy density decreases for a certain time.

\subsection{Non-flat solutions}

In addition to the analytical special solutions discussed above, we would
like to present a few suggestive numerical results
concerning the $k=\pm 1$ cases.
For $\nu=1$, we substitute (\ref{radioactive}), true for any $A$, into the
second order equation (\ref{la4}), to obtain
\be
\frac{\ddot{a}}{a}+\frac{\dot{a}^{2}}{a^{2}}+\frac{k}{a^{2}}+(1+3w)
\beta\rho_{1}^{2}\,\frac{e^{-2At}}{a^{6(1+w)}}-(1-3w)\gamma\rho_{1}\,
\frac{e^{-At}}{a^{3(1+w)}}=0\,.
\label{back}
\ee
It is obvious from (\ref{back}) that for outflow with $k=1$ and $w\geq -1/3$
the Universe will exhibit eternal deceleration.
In particular, in the case of dust, Figure 1
depicts the solution for $a(t)$ of (\ref{back}) for some initial conditions for
$a, \rho$. Notice that $a(t)$ after a period of decrease, starts increasing again, thus
delaying its re-collapse.
Of course, with appropriate initial conditions one also obtains solutions
with the standard FRW behavior.

\EPSFIGURE[h!]{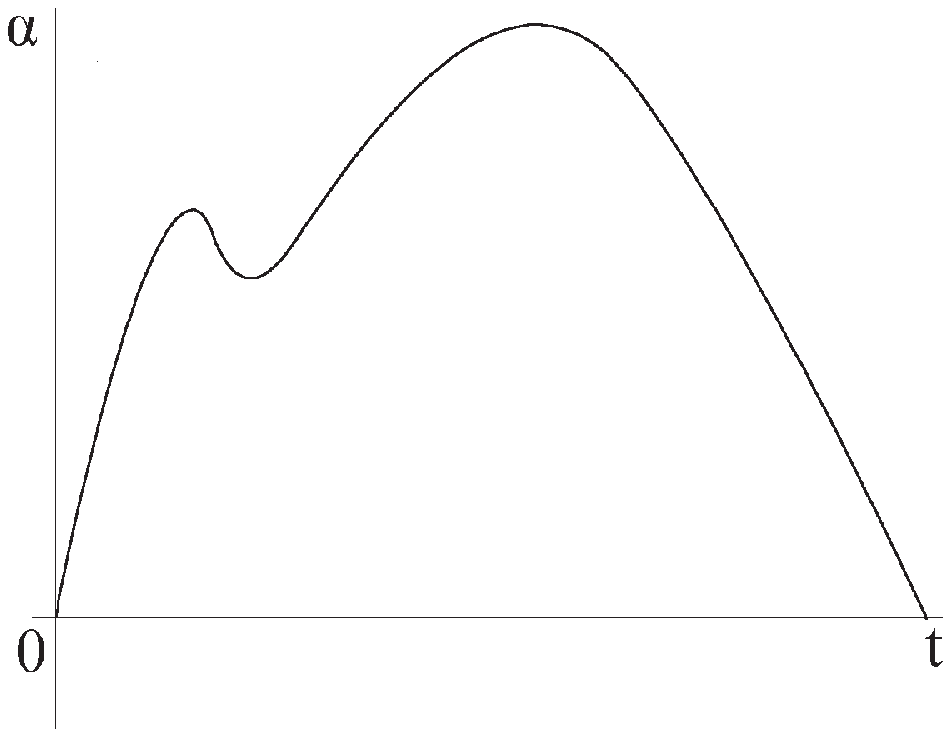,width=200pt,height=150pt}{Outflow,
$k=+1$, $w=0$, $\nu=1$.}

\par
For an open ($k=-1$) Universe with $A>0,\,w=0,\,\nu=1$ we have found
numerically a solution where $\rho(t)$ monotonically decreases to zero, while
$a(t)$ starts with deceleration, but later on accelerates eternally;
more specifically, $q(t) \rightarrow 0^{+}$ for $t \rightarrow 0$. Another
possibility for $k=-1$ and $A<0,\,w=0,\,\nu=1$ allows for a Universe starting
with acceleration at infinite densities, which later on, turns to
deceleration with $\rho$ approaching a constant value.

\section{General treatment of the outflow/influx equations}

In this section, we present a few first steps towards an exact
analytical treatment of equations (\ref{rho})--(\ref{chi})
that govern the cosmological evolution. As we mentioned earlier, we
concentrate on the case with zero effective cosmological constant
$\lambda$ and investigate the brane cosmological
dynamics due to the energy exchange with the bulk.

Some qualitative properties of the exact solutions can be derived from
general arguments:

a) The evolution equations for $T=0$ reduce to those of
ref. \cite{binetruy}: (\ref{rho}) is
the standard energy conservation equation,
while (\ref{chi}) upon integration leads to
the usual mirage radiation term $\chi=C/a^4$.
If there is no energy exchange between the brane and the bulk
($T=0$), it can be seen from equation (\ref{decel})
that for suitable $C<0$ an acceleration
era ($\ddot a>0$)
is possible. It is tempting to associate this
era with the recently observed cosmic acceleration \cite{perl}. It is an open
question, however, if the preceding deceleration era (with a dominant
contribution $\sim \rho^2$ to the Hubble parameter) can accommodate the
successes of standard Big-Bang cosmology.

b) For $T\not= 0$, a
suitable negative $\chi$ term can, in principle, produce acceleration (i.e.
behave like a positive cosmological constant in equation (\ref{decel}))
and at the same time play a role
similar to that of a negative cosmological constant in the Hubble expansion
(\ref{a}). A consequence of this fact is that, if equations
(\ref{rho})--(\ref{decel}) are to describe the present-day Universe
with ${\ddot{a}}/{a}>0$, we must require $\chi <0$. In the asymptotic regime
$\rho \ll {\gamma}/{\beta}$, this means that, for $k=0$, the matter content of
the critical Universe must compensate for the negative contribution from
$\chi$. This requires $\Omega_m > 2$ and $\Omega_\chi < -1$.
As a result, this scenario
predicts more dark energy than
conventional FRWL models. Notice, however, that
for $k=-1$ these restrictions do not apply.
For example, we could have $\Omega_{k}=0.96$,
$\Omega_{\chi}=-0.06$ and $\Omega_{m}=0.1$.

c) In the case of outflow ($T>0$), it is obvious from the conservation equation
(\ref{rho}) that $\rho(t)$ is monotonically decreasing for all expanding
solutions. Furthermore, if the function $T(\rho)$ vanishes only at $\rho
=0$, $\rho(t) \to 0$ for $t\to +\infty$.

d) It is straightforward to argue that for $k=0,+1$ and $ w>-1/3 $, when $\rho
> (1-3w)\gamma/((1+3w)\beta)$ we have necessarily deceleration. For $k=0,+1$,
$w\geq 1/3$ we have deceleration at all times. The absolute upper bound for
$\ddot a/a$ is at all times $(1-3w)^2\gamma^2/(4(1+3w)\beta)$. These
conclusions follow from the fact that the left hand side of equation (\ref{a}) must be
positive definite. This implies that ${\ddot a}/a$ lies below the parabola
$(1-3w)\gamma\rho-(1+3w)\beta\rho^2$, which for $w\geq 1/3$ is negative
definite. Finally, one may see from equations (\ref{a}), (\ref{chi}) that all
expanding $k=0$ solutions with $\rho\to 0$ for $t \to +\infty$, will have
$\chi\to 0$, $\ddot a/a \to 0^{-}$ and $\dot a/a \to 0^+$, for functions
$T(\rho)$ with $T(0) = 0$, while for $T(0) < 0$, $\rho \rightarrow 0$ is not
possible.

\subsection{General analysis of the equations}
Combining equations (\ref{rho}) and (\ref{a}) we obtain
\be
a{{d\rho}\over da}=-3(1+w)\rho-
\epsilon\,T(\rho) \left(\beta\rho^2+2\gamma \rho
- {k\over{a^2}}+\chi\right)^{-1/2}.
\label{rho(a)}
\ee
Similarly, equations (\ref{chi}) and (\ref{a}) give
\be
a{{d\chi}\over da}=-4\chi+2\epsilon \beta\left(\rho+{\gamma\over\beta}\right)
T(\rho)
\left(\beta\rho^2+2\gamma \rho - {k\over{a^2}}+\chi\right)^{-1/2},
\label{chi(a)}
\ee
where $\epsilon =1$ refers to expansion, while $\epsilon=-1$ to contraction.
These two equations form a two-dimensional dynamical system. The function
$\chi(\rho)$ is obtained from the equation
\begin{eqnarray}
\Biggl(3(1+w)\rho
\sqrt{\beta\rho^2+2\gamma \rho -{k\over{a^2}}+\chi}&+&\epsilon\,T(\rho)
\Biggr)\,
{{d\chi}\over{d\rho}}
\nonumber \\
=4\chi\sqrt{\beta\rho^2+2\gamma \rho-{k\over{a^2}}+\chi}
&-&2\epsilon \beta\left(\rho+{\gamma\over\beta}\right)T(\rho).
\label{chi(rho)}
\end{eqnarray}
Note that the equations of contraction are those of expansion with the
roles of outflow and influx interchanged. Combining equations (\ref{chi(rho)})
with (\ref{decel}) we derive the following equation for $q(\rho)$:
\begin{eqnarray}
\Biggl(&&3(1+w)\rho
+\epsilon T(\rho)/\sqrt{(1-3w)\gamma\rho-(1+3w)\beta\rho^2-ka^{-2}-q(\rho)}\,
\Biggr)
\frac{dq(\rho)}{d\rho} \nonumber \\
&&+\epsilon \Bigl(2(1+3w)\beta\rho-(1-3w)\gamma\Bigr)T(\rho)/
\sqrt{(1-3w)\gamma\rho-(1+3w)\beta\rho^2-ka^{-2}-q(\rho)} \nonumber \\
&-&4q(\rho)+
(1+3w)\Bigl(2(2+3w)\beta\rho-(1-3w)\gamma\Bigr)\rho=0\,.
\label{q}
\end{eqnarray}
We shall use this equation below to decide about eras of
acceleration and deceleration in the cosmic evolution on the brane. For
$|w|<1/3$, $T(\rho)>0$ and $\rho<((1-3w)\gamma)/(2(2+3w)\beta)$, it can be seen
from the above equation that $q>0$ implies $dq/d\rho>0$. However, in the
range $((1-3w)\gamma)/(2(1+3w)\beta) < \rho < ((1-3w)\gamma)/((1+3w)\beta)$,
passage from acceleration to deceleration during expansion is not allowed.

Another form of the system of equations (\ref{rho(a)}), (\ref{chi(a)}),
which will be helpful for the study of fixed points of the system, can
be derived by defining $\alpha\equiv \ln{a}$ and $Q\equiv q+k e^{-2\alpha}$.
Then
\be
{{d\rho}\over {d\alpha}}=R(\rho,Q)~~~~~~,~~~~~~{{dQ}\over
{d\alpha}}=2ke^{-2\alpha}+S(\rho,Q),
\label{rhoQ}
\ee
where
\begin{eqnarray}
R(\rho,Q)&=&-3(1+w)\rho-\epsilon T\Bigl(-(1+3w)\beta\rho^2+(1-3w)\gamma
\rho-Q\Bigr)^{-1/2},
\label{R}  \\
S(\rho,Q)&=&(1+3w)(2(2+3w)\beta\rho-(1-3w)\gamma)\rho-4Q \nonumber \\
&&+\epsilon T\Bigl(2(1+3w)\beta\rho-(1-3w)\gamma
\Bigr)\Bigl(-(1+3w)\beta\rho^2+(1-3w)\gamma \rho-Q\Bigr)^{-1/2}.
\nonumber \\
&&\label{S}
\end{eqnarray}
\par
When the 3-metric is flat ($k=0$), equation
(\ref{chi(rho)}) becomes an autonomous equation, which, in principle, may be
solved for $\chi(\rho)$. We define
\be
Y(\rho)\equiv \rho^{-{2\over{3(1+w)}}}\sqrt{\beta\rho^2+2\gamma\rho+\chi}
+{\epsilon\over{3(1+w)}}\rho^{-{{5+3w}\over{3(1+w)}}}T(\rho),
\label{Y}
\ee
in terms of which equation (\ref{chi(rho)}) translates to
\begin{eqnarray}
Y{{dY}\over{d\rho}}&&+{\epsilon\over{3(1+w)}}\rho^{-{{8+6w}\over{3(1+w)}}}
\Biggl({{7+3w}\over{3(1+w)}}T-\rho{{dT}\over{d\rho}}\Biggr)Y
-{2\over{27(1+w)^3}}\rho^{-{{13+9w}\over{3(1+w)}}}T^2 \nonumber \\
&&+\left({{1-3w}\over{3(1+w)}}\gamma-{{1+3w}\over{3(1+w)}}\beta\rho\right)
\rho^{-{4\over{3(1+w)}}}=0.
\label{universe}
\end{eqnarray}

We shall simplify our discussion by considering energy transfers of the form
\be
\epsilon\,T(\rho)=A\rho^\nu,
\label{T}
\ee
with $\nu$ an arbitrary real parameter. The brane-bulk energy
exchange in the real world may be much more complicated.
For example, it may be a sum of terms of the above form.

With the change of variable
\be
\bar\rho=\rho^{\nu-{{5+3w}\over{3(1+w)}}},
\label{rhobar}
\ee
the equation
for $Y$ becomes
\begin{eqnarray}
Y{{dY}\over{d\bar\rho}}&-& BY
\nonumber \\
&+&{1\over{5+3w-3\nu(1+w)}}
\left({{2A^2}\over{9(1+w)^2}}\bar\rho+(1+3w)\beta\bar\rho^{-r}
-(1-3w)\gamma\bar\rho^{-s}\right)=0, \nonumber \\
\label{universe2}
\end{eqnarray}
where $B\equiv
A(7+3w-3\nu(1+w))/(3(1+w)(5+3w-3\nu(1+w)))$, $r\equiv
(7+9w-3\nu(1+w))/(5+3w-3\nu(1+w))$ and $s\equiv
(4+6w-3\nu(1+w))/(5+3w-3\nu(1+w))$.

In the cases with $-1 < w \leq 1/3$, $\nu \leq (4+6w)/(3(1+w))$,
or $w \geq 1/3$, $\nu < (5+3w)/(3(1+w))$, or $-1 < w \leq -1/3$, $\nu >
(5+3w)/(3(1+w))$, or $w \geq -1/3$, $\nu \geq (7+9w)/(3(1+w))$, we have $r,s
\geq 0$. In the case $w \geq 1/3$, $(5+3w)/(3(1+w)) < \nu \leq 3/2$, we have
$r,s \leq -1$. In all these cases, the behavior $\rho \approx
0$, if such a region exists, is given by the following approximation of equation
(\ref{universe2}):
\be
Y \frac{dY}{d
\bar{\rho}}-BY+\frac{2A^2}{9(1+w)^{2}(5+3w-3\nu (1+w))} \bar{\rho}=0.
\label{asymptotic}
\ee

\subsection{Solutions with $\rho \approx 0$}

{\it Case 1}: For $\nu =1$ (and thus $w > -1/3$), as in the case of
unstable matter on the brane decaying into the bulk, the general solution of
(\ref{asymptotic}) is \cite{kamke}
\be
\left(\bar\rho-\frac{3(1+w)}{A}Y
\right)
\exp\left({{{\bar\rho}\over{\bar\rho-\frac{3(1+w)}{A}Y}}}\right)
=-\frac{3(1+w)}{A}\kappa.
\label{papa}
\ee
Equivalently, in terms of the function $\chi$,
the above solution takes the form
\be
\sqrt{\beta\rho^2+2\gamma\rho+\chi}\,\,\,
\exp\left(-{A\over{3(1+w)}}{1\over{\sqrt{\beta\rho^2+2\gamma\rho+\chi}}}
\right)
=\kappa \rho^{2\over{3(1+w)}},
\label{solution}
\ee
where $\kappa$ is a non-negative integration constant.
Using (\ref{a}) we may rewrite the above as
\be
|H| e^{-{A\over{3(1+w)}}{1\over |H|}}=\kappa \rho^{2\over{3(1+w)}}.
\label{rhoofhubble}
\ee

Equation (\ref{rhoofhubble}), as is suggested by its independence on
$\beta$ and $\gamma$,
is valid only as long as the $\beta$ and $\gamma$ terms in (\ref{universe2})
are negligible compared to the linear term inside the parenthesis. In the case
of outflow this is not a constraint, since $\rho$ is monotonically decreasing
towards zero. So, at late times all solutions are correctly described by
(\ref{rhoofhubble}).

From equation (\ref{papa}) we conclude that for $A<0$ (influx), $\nu=1,\,w >
-1/3$, any solution $\rho(t)$ with $\kappa > 0$ is bounded from below by some
positive value. This implies that the point $(\rho=0, q=0)$ cannot be an attractor.
Furthermore, using (\ref{rhoofhubble}) and (\ref{la4}) one may show that for $w\neq 1/3$
$q\simeq (1-3w)\gamma\rho$, i.e. all curves near the origin of the $(q,\rho)$ plane
have a fixed slope. This will be confirmed numerically in Figure 3 below.
Also, $|H|\ll \rho/M^3$.

In the case of outflow, on the other hand, equation (\ref{rhoofhubble}) implies
that for $\nu=1,\,w > -1/3$, we have $|H|\to 0$ for $\rho \to 0$.
Equation (\ref{rhoofhubble}) implies that $\rho\ll |q|$ and in addition that
$\rho/q\to -\infty$ as $\rho\to 0$, in agreement with the case depicted in Figure 2.
Using (\ref{rho})--(\ref{chi}) it is straightforward to show that
\be
{{dH^2}\over {d\alpha}} + 4H^2-2(1-3w)\gamma\rho+2(1+3w)\beta\rho^2=0,
\label{last}
\ee
which combined with (\ref{rhoofhubble}) leads to
\be
|H|={{c_1}\over{a^2}}\, ,
\label{h(t)}
\ee
or equivalently to
\be
a^2(t)=2c_1 t+c_2,
\label{a(t)}
\ee
with $c_1$ and $c_2$ integration constants.
Using this expression of $H(t)$ in (\ref{rhoofhubble}), we obtain
\be
\rho(t)=(2\kappa)^{-3(1+w)/2}\frac{e^{-A(t+{{c_2}\over{2c_1}})}}{(t+{{c_2}\over{2c_1}})^{3(1+w)/2}}.
\ee
Apart from the obvious agreement of the above with the special solution
(\ref{solrad1}) (also valid for $A>0$),
we should like to stress at this point the remarkable fact that the scale factor $a(t)$ of the Brane
Universe, for $\nu=1$ and essentially any $w$, behaves at late times as if radiation dominated.

\EPSFIGURE[h!]{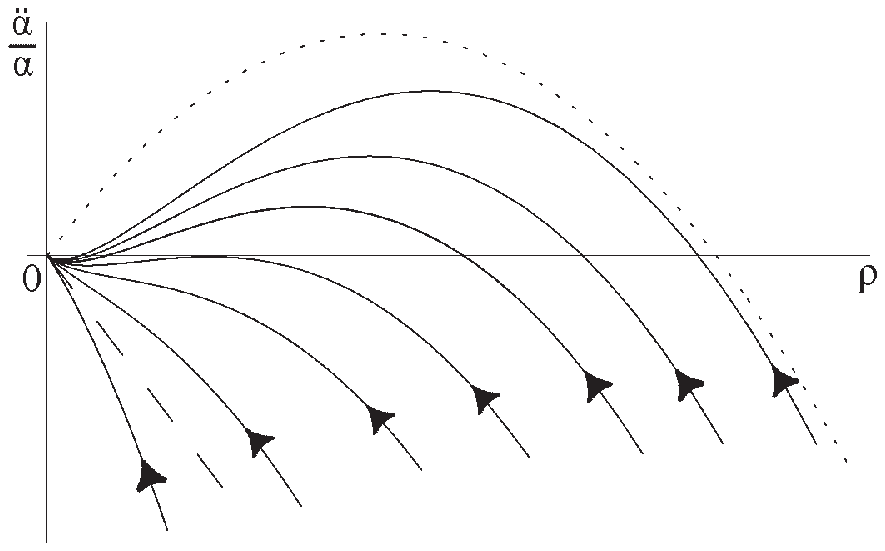,width=200pt,height=150pt}{Outflow, $k=0$, $w=0$,
$\nu=1$. The arrows show the direction of increasing scale factor.}

A global phase portrait of $q\equiv\ddot{a}/a$ with respect to
$\rho$ during expansion in the outflow case for $k=0,\,w=0, \,\nu=1$ is shown
in Figure 2. The characteristics of the solutions are in agreement with our
conclusions above.
One recognizes two families of solutions:
The first have $q<0$ for all values of $\rho$, while the  second
start with a deceleration era for large $\rho$, enter an acceleration era and then
return to deceleration for small enough values of $\rho$.
This is exactly what should be expected on the basis of the previous analysis.
Solutions corresponding to initial conditions with positive $q$
(always under the limiting parabola shown with the dotted line),
necessarily had a deceleration era in the past,
and are going to end with an eternal deceleration era also.
The
straight dashed line represents the standard FRW solution.

{\it Case 2}: For $\nu \neq 1$, following the same steps as in case 1, the
general solution of (\ref{asymptotic}) for $-1 < w \leq 1/3$, $\nu \leq
(4+6w)/(3(1+w))$, or $w \geq 1/3$, $\nu < (5+3w)/(3(1+w))$ is
\be
\kappa_1 \left|Y-\frac{A}{3(1+w)}\bar{\rho}\right|\,\,^{^{^{\frac{5+3w-3\nu
(1+w)}{2}}}}= \kappa_2 \left |Y-{{2A} \over {3(1+w)(5+3w-3\nu
(1+w))}}\bar{\rho} \right|,
\label{sola2}
\ee
where $\left|\kappa_1 \right|+\left|\kappa_2\right|>0$. Using (\ref{Y}) and
(\ref{a}) this is translated into the implicit solution for $H(\rho)$
\be
\kappa_{1}|H|\,\,^{^{\frac{5+3w-3\nu (1+w)}{2}}} =\kappa_{2}\left|\rho^{1-\nu}\,
|H|-{{A\,(\nu-1)} \over {5+3w-3\nu (1+w)}}\right|\,.
\label{rhohubble}
\ee

A few comments concerning the validity of our approximations are in order at this point. First, we
would like to stress that (\ref{papa}) and (\ref{sola2}) were derived and are
valid only near "vanishing" $\rho$. Their validity may be questionable in
situations with $\rho$ bounded from below. A detailed perturbative analysis of the
next order correction to $Y(\rho)$, ensures that the above expressions apply in
all cases studied, except for some sets of initial data in the case of energy
influx domination ($A<0$, $\nu<1$),
in which $\rho$ is indeed bounded from below.
A similar comment applies to equations (\ref{rhoofhubble}) and (\ref{rhohubble}).
The translation from $Y$ to $H$ involves equation (\ref{Y}), whose possible singular
nature leads to some further constraints on their validity. Specifically, in the case of
$\nu\geq 1$ for both influx and outflow, (\ref{rhoofhubble}) and (\ref{rhohubble}) are correct
for all solutions with $\rho$ approaching arbitrarily close to zero.
For dominant outflow or influx, i.e. $\nu <1$,
(\ref{rhohubble}) should not be trusted.
In these cases, the next order correction to $H$ is of the same order of magnitude as the
leading $H$, given in (\ref{rhohubble}).
Incidentally, the above clarifications remove an apparent contradiction of
(\ref{rhohubble}) with the general comment (d) in the beginning of this section. Namely,
(\ref{rhohubble}) implies that for $\nu<1$, $H(\rho)$ tends to a non-vanishing
constant as $\rho\to 0$. This contradicts comment (d) but happens in a region of
parameters and initial data for which (\ref{rhohubble}) is not reliable.

Finally, for $1<\nu<3/2$ it is straightforward to see that $q/\rho \to -\infty$
and, furthermore, in the case of outflow,
the explicit expression for the scale factor is given by (\ref{a(t)}),
while the matter density is
\be
\rho(t)=\Biggl(\frac{c_1(5+3w-3\nu(1+w))}{A(\nu-1)(2c_1t+c_2)}\Biggr)^{1/(\nu-1)}.
\ee
In the case of influx
the point $(0,0)$ in the $(\rho-q)$ plane is a repulsor.

\subsection{Fixed points}

Equations (\ref{la3}) and (\ref{la4}) possess for $k=0$ the obvious fixed point solution
$(\rho_*=0,H_*=0)$.
However, there are more fixed points which may be found by setting
${d\rho}/{d\alpha}={dQ}/{d\alpha}=0$ in the system of equations
(\ref{rhoQ}). This leads to the following relations
\be
(1+3w)\beta\rho_{*}^2-(1-3w)\gamma\rho_{*}+{{2|T(\rho_{*})|^2}\over{9(1+w)^2
\rho_{*}^2}}=0,
\label{rhostar}
\ee
\be
q_{*}={{|T(\rho_{*})|^2}\over{9(1+w)^2 \rho_{*}^2}}>0.
\label{qstar}
\ee
It is obvious from the conservation equation (\ref{rho(a)}) that during
expansion non-trivial fixed points may exist only in the influx case. Their number is
determined by the roots of equation (\ref{rhostar}). It can be seen from
(\ref{qstar}) that these solutions
are accelerating. From
(\ref{a}), (\ref{chi}) we find
that $H_{*}=\sqrt{q_{*}}$ and $\chi_{*}=-3(1+w)(\beta
\rho_{*}+\gamma) \rho_{*}/2$.

In order to study the stability of these fixed points
we write $\rho(\alpha)=\rho_{*}+\delta \rho(\alpha)$, $q(\alpha)=q_{*}+\delta
q(\alpha)$. The resulting linearized equations are
\be
{d\over{d\alpha}}
\left(\begin{array}{c}
\delta \rho(\alpha) \\ \delta q(\alpha)
\end{array}
\right)=
{{9(1+w)^2 \rho_{*}^3}\over{2T_{*}^2}}\,\,
\left(\begin{array}{cc}
 m_1 & m_2 \\
m_3 & m_4
\end{array}\right)
\left(\begin{array}{c}
\delta \rho(\alpha) \\ \delta q(\alpha)
\end{array}
\right),
\label{matrix}
\ee
where
\begin{eqnarray}
m_1&=&3(1+w)\,((3-\nu)(1+3w)\beta \rho_{*}+(\nu-2)(1-3w)\gamma)
\nonumber \\
m_2&=&3(1+w)
\nonumber \\
m_3&=&((1-3w)\gamma-2(1+3w)\beta \rho_{*})
\nonumber \\
&&\times\left[(1+3w)(7+9w-3\nu
(1+w))\beta \rho_{*}
-(1-3w)(4+6w-3\nu (1+w))\gamma\right]
\nonumber \\
m_4&=&-(2(1+3w)^2 \beta \rho_{*}+(1-3w)^2 \gamma).
\label{ms}
\end{eqnarray}
The sign of the real part of the eigenvalues of the
matrix appearing above determines the nature of
the fixed point. The presence of an imaginary part results in a spiral
form for the flows.

As an example, we consider the case with dust ($w=0$) and
$\nu=1$. For $|A| \leq {{3\gamma}/{\sqrt{8\beta}}}$, there exist two
real positive roots
\be
\rho^{(\pm)}_{*}=\left(1 \pm \sqrt{1-{{8\beta A^2}\over{9
\gamma^2}}}\right) {{\gamma}\over{2\beta}}.
\label{roots}
\ee
The root $\rho_{*}^{(+)}$  is
always a saddle point. Furthermore, for
${{27 \gamma^2}/({32\beta}}) < A^2 <
{{9 \gamma^2}/({8\beta}})$, $\rho_{*}^{(-)}$ is a stable node, while for
$A^2 < {{27 \gamma^2}/({32\beta}})$, it becomes a
counterclockwise stable spiral. For $\beta A^2/\gamma^2\ll 1$ we
have $\rho_{*}^{(-)}\simeq 2A^2/(9\gamma)\ll \gamma/\beta$ and
$\rho_{*}^{(+)}\simeq \gamma/\beta$. In this limit, $\rho_{*}^{(-)}$
corresponds to the fixed point (\ref{sol1}) of the special solutions
discussed in section 3. The eigenvalues of the stability matrix are in
agreement with the ones derived there.

The global phase portrait of $q\equiv \ddot{a}/a$ with respect to $\rho$ during
expansion for the case $k=0$, $w=0,\nu=1$ is shown in Figure
3. The presence of the limiting parabola as in the outflow case is apparent.
However, new
characteristics appear. For example, $\rho_*^{(-)}$ attracts to eternal
acceleration a whole family of solutions which start their evolution at either
very low or very high densities. There is another family of solutions which
are attracted to acceleration by $\rho_{*}^{(+)}$ and which eventually exit to
a deceleration era.
Finally, there is a family of solutions, near the limiting parabola, which
start with acceleration at very low
densities, and eventually exit to eternal
deceleration, while their density increases
monotonically with time because of the influx.

\EPSFIGURE[h!]{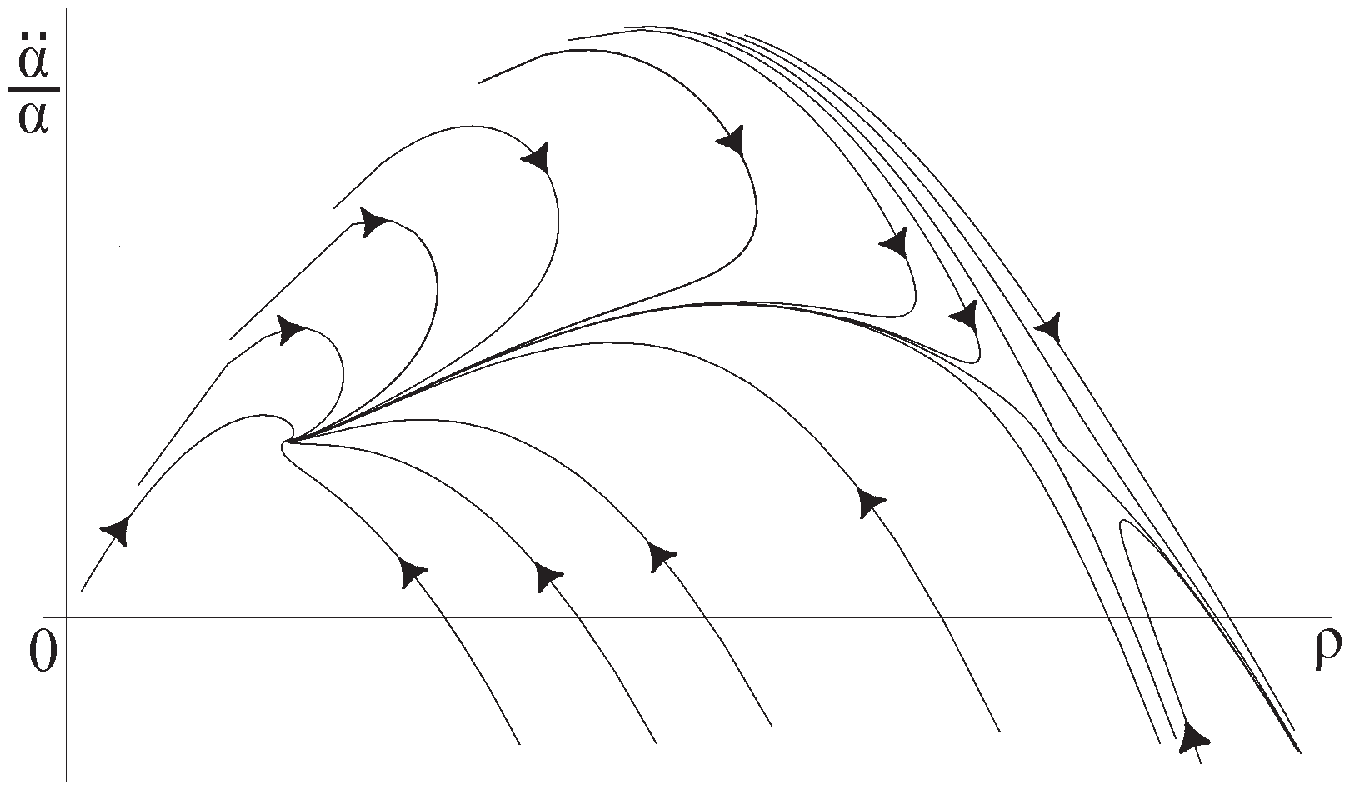,width=200pt,height=150pt}{Influx, $k=0$, $w=0$,
$\nu=1$.}

Clearly, for $\nu\neq 1$ one expects a different set of fixed points with
varying behaviors around them.

\section{Conclusions}

A rather detailed mathematical analysis of the role of the brane-bulk energy
exchange on the evolution of a Brane
Universe, adequate for a wide range of potentially realistic implementations,
was presented.
The effective brane cosmological equations were derived with perfect fluid matter
on the brane, constant energy-momentum tensor in the bulk and non-vanishing
exchange between them.
A detailed study of the solutions of these unconventional equations was performed
in the case of zero effective cosmological constant on the brane, in order to reduce
to the Randall-Sundrum vacuum in the absence of matter.
The analysis revealed a
rather rich variety of possible cosmologies, depending upon the precise form of
the exchange term, the topology of 3-space, and the nature of matter on the brane.

A few special but particularly interesting solutions were obtained in the limit
of low energy density on the brane.
One of these is the exactly
solvable case of radiation, where a mirage radiation effect appears through the
decay of real brane matter. Another, is a De Sitter fixed point solution,
obtained in the case of energy influx, even without pure vacuum energy, which in addition
is stable for a wide range of reasonable forms of energy exchange.

An exact treatment of the full set of equations followed. Several generic
qualitative bounds were extracted and the asymptotic characteristics of the solutions
were studied. In particular, in the low-density regime and for a flat
universe, our model predicts additional dark matter compared to the standard cosmological
picture.
Adopting the physically motivated $T\sim \rho^\nu$ power-law form for the energy transfer,
we found exact analytical solutions in the low-density regime for almost all values of $\nu$
and of the matter equation of state. Furthermore, the global phase portraits in the
density-acceleration plane were obtained numerically, separately for the outflow and the influx cases,
for arbitrary values of the
density.
In the outflow case, one distinguishes two families of solutions: one describing
an all-time decelerating universe, and another which describes a Universe with an
intermediate accelerating era. The latter starts decelerating, enters an accelerating
phase, and finally decelerates again, with the energy density decaying to zero.
In the case of influx on the other hand, one generically obtains several fixed point solutions,
whose stability analysis was performed. Curiously, all these fixed points
correspond to positive acceleration, and their presence implies the existence
of solutions with accelerating era. In addition, as one may see in Figure 3, there are
solutions that can be purely
accelerating, purely decelerating, or with alternating accelerating
and decelerating behavior.

A complete description of the evolution of our Brane Universe is still lacking.
Several scenaria are possible and their detailed
study in the light of observations will determine their viability.
For example, a sketchy cosmological evolution could be as follows:
The brane is created much hotter
than the bulk with a very large energy density somewhere
near the limiting parabola. It emits energy to the bulk,
and after the initial decelerating period, it acquires an acceleration
(presumably, the primordial "inflation"),
which for appropriate choice of parameters and initial conditions can be
of the order of $\gamma^2/\beta$.
It quickly cools down and starts decelerating, with energy density much smaller than
the one it had initially.
Assuming that the energy emission is very fast, it happens with the brane-bulk
system out of equilibrium, and one may expect that the brane will supercool and reach
a temperature much smaller than the roughly constant temperature of the bulk.
As a consequence of the energy influx, the evolution of the brane will be attracted
to the stable fixed point, analogous to the $\rho_*^{(-)}$
discussed in the text. The corresponding $q_* \sim \gamma \rho_*^{(-)}$ will naturally be much smaller
than its early value, and may fit today's cosmic acceleration \cite{perl}.

Clearly, the details of scenaria such as the above
require further study and many open questions, such as the duration of
accelerating periods, the creation of primordial fluctuations,
and the compatibility with conventional cosmology at low energy densities,
should be addressed. However, we believe that the
cosmological evolution in the context we presented here has
many novel features, that may provide answers to outstanding
questions of modern cosmology.

Some of the cosmological evolutions presented will have a (more general) holographic dual
that will be interesting to understand in more detail. This will provide a more controllable picture
of the mechanisms advocated in this paper.

\section*{Acknowledgments}

This work was partially
supported by European Union under the RTN
contracts HPRN--CT--2000--00122 and --00131.
The work of E. K. was supported by Marie Curie
contract MCFI-2001-0214.
We would like to thank the referee for helping clarify some points in this paper.

\end{document}